# Inertial Migration of Aerosols in Microfluidic channels

Maoqiang Jiang[a], Shizhi Qian[b], Zhaohui Liu[a]†

[a]*State Key Laboratory of Coal Combustion, School of Energy and Power Engineering, Huazhong University of Science and Technology, Wuhan 430074, China*

[b]*Department of Mechanical and Aerospace Engineering, Old Dominion University, Norfolk, VA 23529, USA*

† Corresponding author. Tel.: +86-027-8754-2417-8308. E-mail address: zliu@hust.edu.cn (Z. Liu).

**Abstract**

In recent years, manipulation of particles by inertial microfluidics has attracted significant attention. Most studies focused on inertial focusing of particles suspended within liquid phase, in which the ratio of the density of the particle to that of the medium is $O(1)$. The investigation on manipulation of aerosol particles in an inertial microfluidics is very limited. In this study, we numerically investigate the aerosol particle motion in a 3D straight microchannel with rectangular cross section by fully resolved simulation of the particle-air flow based on the continuum model. The air flow is modeled by the Navier-Stokes equations, and particle's motions, including translation and rotation, are governed, respectively, by the Newton's second law and the Euler equations without using any approximation models for the lift and drag forces. The coupled mathematical model is numerically solved by combining immersed boundary with lattice Boltzmann method (IB-LBM). We find that the Reynolds number (Re), the particle's initial position, particle's density, and particle's diameter are the influential parameters in this process. The equilibrium positions and their stabilities of aerosols are different from those suspended in liquid.

## 1. Introduction

Air pollution with suspended aerosols or airborne particulate matters (PM) has been extremely serious today, either indoor or outdoor. One of the most hazardous matter is bioaerosol, such as viruses, bacteria, pollen, fungi and enzymes, etc. Once these airborne pathogenic microorganisms contact with human body or domestic animals, they can cause serious diseases, including asthma, allergies, pneumonia and infectious diseases (Douwes et al. 2003). Therefore, it is important and crucial to detect them rapidly and effectively for disease control or human health protection (Zhang et al. 2018). To monitor these pathogenic airborne microorganisms (Hong et al. 2015), the first key step is to sample bioaerosols. However, the bioaerosols have several kinds, with their sizes ranging from 20 nanometer to 200 micron in diameter. In addition, they are suspended along with other particles, such as dusts, droplets, and salts, et al. Therefore, efficient separation, collection or capture of the target aerosols from background environment is the foremost step prior to subsequent bioanalysis (Bian et al. 2016). Traditional methods for isolation and identification of the target aerosols, such as the semiquantitative plate sedimentation culture method (Yu et al. 2017), mass spectrometry (MS) (Pietrowska et al. 2009), and loop-mediated isothermal amplification (LAMP) (Han 2013) are time-consuming, difficult and complicated (Zhang et al. 2018).

Microfluidics, a new technique rapidly developed since 1990s, becomes another prominent method to provide direct capture and downstream airborne pathogen analysis (Metcalf et al. 2018; Sui and Cheng 2014). Over the past few years, several studies have demonstrated separation of submicron aerosols and bioaerosols on microchips according to their inertial differences, size differentiation and special structure (Schaap et al. 2012). It is simple, rapid, portable and low-cost. For example, Chin et al. (Chin et al. 2011) used microfluidics to enable a single, easy to use point-of-care (POC) diagnostics and early detection of infectious diseases. Jing et al. (Jing et al. 2014; Jing et al. 2013) successfully developed a simple microfluidic device for capture and enrichment of airborne bacteria by using S-shaped microchannel and staggered herringbone mixer structures. Hong et al. (Hong et al. 2015) presented a curved microchannel to separate viruses and bacteria form large particles by centrifugal force. Recently, a double-spiral sawtooth wave-shaped microchannel with herringbone structures was designed by Bian et al. (Bian et al. 2016). It can be simply fabricated and is fast for efficient capture of bacterial from polluted air. Liu et al. (Liu et al. 2016) fabricate a first portable direct analysis system for the rapid detection of airborne pathogen using a microfluidic chip which has both enrichment and identification functions. Yin et al. (Yin et al. 2017) demonstrated an I-shape pillar based deterministic lateral displacement (DLD) microfluidic device for portable and real-time monitoring of PM2.5.

However, the aforementioned designs mainly depend on experience. The particle's detailed motion in microchannels is unclear, though a large body of research has been dedicated to particles' motions suspended in liquid instead of air. In a straight liquid microchannel, initial randomly distributed neutrally buoyant particles will finally migrate to several symmetry equilibrium positions due to the inertial effect (Liu et al. 2015; Martel and Toner 2014; Zhang et al. 2016). However, the aerosol particles are no longer neutrally buoyant. And the properties of air medium are much different from those of liquid medium, especially the density and viscosity. The dynamics and transport process of aerosol particles are expected to be different from those of particles immersed in liquid medium. Motivated by this, we investigate the detailed dynamics of aerosol particles in microfluidic channels by fully resolved simulation based on immersed boundary-lattice Boltzmann method (IB-LBM) without using any empirical formula for forces acting on the particle.

## 2. Theoretical analysis

The air flow in a microchannel can be described by the Knudsen number, $Kn = \lambda / L$, which is the ratio of the mean free path of air, $\lambda$, to the representative geometry length scale, $L$. Here, $L$ is the hydraulic diameter of the 3D microchannel with a rectangular cross-section, $D_h = 2HW / (H + W)$, with $H$ and $W$ being height and width of the microchannel, respectively. The calculated $Kn \ll 1.0$ according to the physical conditions in this study, indicating that the flow is continuum in the current study.

Next, we discuss the differences between the air-particle system and the conventional liquid-particle system. As shown in Table 1, the physical parameters of the two systems, such as particle density $\rho_p$, particle diameter $D_p$, channel height $H$ and inlet fluid average velocity $U_f$ are set to be the same. The air density $\rho_f$ is only about 1.205 kg/m$^3$, which is much lower than that of liquid with 1000 kg/m$^3$. This yields that the fluid kinematic viscosity in the air-particle system is larger than that in the liquid-particle system, while the fluid dynamic viscosity in the former is much lower than that in the latter. For example, the dynamic viscosity $\mu_f$ and kinematic viscosity $\nu_f$ of air are about $1.79 \times 10^{-5}$ Pa s and $1.48 \times 10^{-5}$ m$^2$ s$^{-1}$, respectively. However, those of liquid (i.e. water) are about $1.01 \times 10^{-3}$ Pa s and $1.01 \times 10^{-6}$ m$^2$ s$^{-1}$, respectively. As a result, the fluid's inertial, i.e. channel Reynolds number, is much smaller in the air-particle system than that in the liquid-particle system. However, the particle's inertial, i.e. Stokes number (Brennen 2005), is much larger in the air-particle system than that in the liquid-particle system. Here, channel Reynolds number $Re_c$ and particle Stokes number $St_p$ are defined as follows (Akhatov et al. 2008b; Hoey et al. 2012):

$$Re_c = \frac{\rho_p U_f H}{\mu_f} \tag{1}$$

$$St_p = \frac{\tau_p}{\tau_f} = \frac{\rho_p D_p^2 / 18\mu_f}{H/U_f} = \frac{\rho_p D_p^2 U_f}{18\mu_f H}, \tag{2}$$

where $\tau_p$ and $\tau_f$ are the relaxation times of the particle motion and fluid flow, respectively. Particles with small Stokes number can easily follow the fluid flow, while particles with large Stokes number can detach from fluid flow when the fluid flow changes magnitude or direction. Large Stokes number will also induce particles to lag behind the fluid in fluid-driven particle motion, and produces a larger slip-shear velocity between particle and fluid (Akhatov et al. 2008a; Deng et al. 2008; Jebakumar et al. 2016). As shown in Table 1, particle Stokes number is only 0.22 at moderate Reynolds number ($Re_c$=100) in a liquid-particle system, while it becomes 12.54 at low Reynolds number ($Re_c$=6.8) in an air-particle system. Therefore, the ratio of the particle Stokes number to fluid Reynolds number in an air-particle system is 1.84, which is about 10$^3$ times of that in a liquid-particle system (i.e., $2.2 \times 10^{-3}$). This suggests that the particle's inertia is relatively more important in the air-particle system.

**Table 1.** Comparison of parameters in a liquid (water)-particle system and an air-particle system in a square microfluidic channel.

|  | Parameters | Liquid-particle system |  | Air-particle system |
|---|---|---|---|---|
| Same | Particle density, $\rho_p$ | 1000 kg m$^{-3}$ | = | 1000 kg m$^{-3}$ |
|  | Particle diameter, $D_p$ | 5 μm | = | 5 μm |
|  | Channel height, $H$ | 25 μm | = | 25 μm |
|  | Inlet fluid average velocity, $U_f$ | 4.04 m s$^{-1}$ | = | 4.04 m s$^{-1}$ |
| Difference | Fluid density, $\rho_f$ | 1000 kg m$^{-3}$ | >> | 1.205 kg m$^{-3}$ |
|  | Fluid dynamic viscosity, $\mu_f$ | $1.01 \times 10^{-3}$ Pa s | >> | $1.79 \times 10^{-5}$ Pa s |
|  | Fluid kinematic viscosity, $\nu_f$ | $1.01 \times 10^{-6}$ m$^2$ s$^{-1}$ | < | $1.48 \times 10^{-5}$ m$^2$ s$^{-1}$ |
|  | Channel Reynolds number, $Re_c$ | 100 | >> | 6.8 |

| Particle Stokes number, $St_p$ | 0.22 | << | 12.54 |

In order to explain particle's motion and behavior, we further analyze forces acting on particles. In the liquid-particle system, there are two dominant forces: shear gradient lift force induced by the parabolic velocity profile and the wall induced lift force caused by the hydrodynamic repulsive effect between particle's surface and channel wall (Amini et al. 2014; Martel and Toner 2014; Zhang et al. 2016). The former force directs towards the wall and the latter towards the centerline, as shown in Fig. 1(a). Particles pushed by these two forces are finally focused to an equilibrium position in the crossflow direction when the two forces are balanced each other. However, in an air-particle system, there are at least two additional forces, as shown in Fig. 1(b). One is the Saffman lift force (Akhatov et al. 2008a; Saffman 1965) induced by the large slip-shear velocity between fluid and particle. This force directs towards the centerline when the particle lags behind the fluid, while towards the wall when it leads the fluid. The other is the buoyant force as a result of the large density ratio of the heavy particle to the air fluid, which is about $O(10^3)$. Its direction is always towards downward because the gravity always directs towards downward. In fact, because the particle has a finite size, the slip-shear velocities between its two ends, one of which is close to the channel centre with a high air velocity and the other is close to the channel wall with a low air velocity, are unequal. This may cause the particle to rotate and generates another lateral force on the particle, called Magnus force (Barkla and Auchterlonie 1971). Its direction changes depending on the rotation direction of the particle and its magnitude depends on the relative angular rate(Hoey et al. 2012). Therefore, there are at least four or more forces dominating the particle motion in an air-particle system. In addition, the smaller dynamic viscosity in the air-particle system can cause the velocity profile to be steeper with much larger shear rate (Bruus 2008), as shown in the left part of Fig.1(b). This will increase the shear gradient lift force. Therefore, the force acting on a particle in an air-particle system is much more complicated and significantly different from that in the liquid-particle system. Note that Akhatov's group (Akhatov et al. 2008a; Akhatov et al. 2008b; Hoey et al. 2012) have done a lot of work to develop the force model of particle motion for the 3D direct-write technologies based on aerosol jet system. However, its model is not appropriate for the conditions in our study. One reason is that the fluid velocity is very high (about 50~200 m/s) in the aerosol direct-write system with a certain compressible effect, while it is low (less than 10 m/s) in the aerosol sampling system with incompressible effect. The other reason is that the finite size of the particle in the former system is neglected, while should be considered in microfluidics as indicated by many previous literatures (Amini et al. 2014; Amini et al. 2012; Di Carlo et al. 2009) and the simulation results depicted in Section 4.

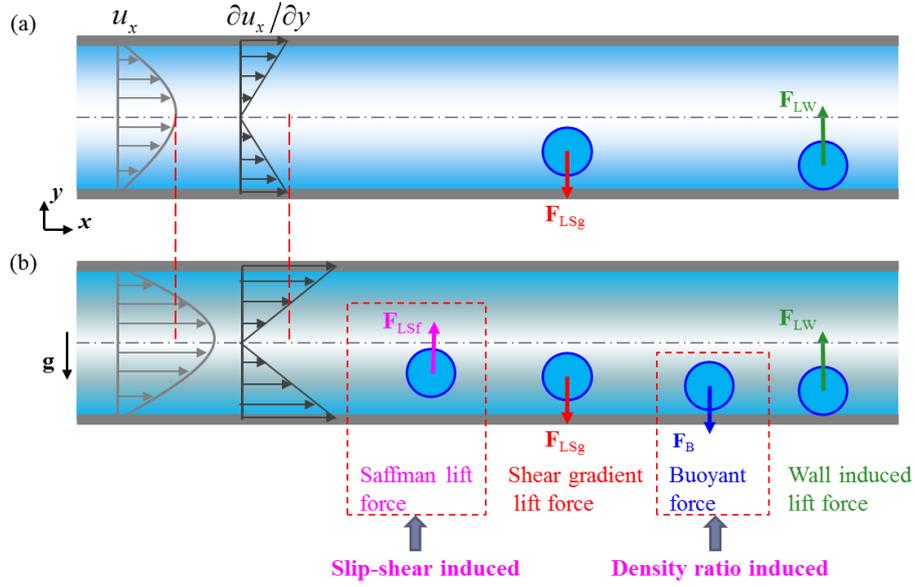

**Fig. 1**. Forces acting on a particle in (a) a liquid-particle system and (b) an air-particle system.

In this work, we will implement the fully resolved direct numerical simulation (FR-DNS) (Maxey 2017) based on immersed boundary-lattice Boltzmann method (IB-LBM) to consider all the forces and the finite-size effect. By doing this, the construction of the complicated force model acting on a particle is avoided, while it is the key in the conventional Euler-Lagrangian method based on a point particle model (Akhatov et al. 2008b; Lee et al. 2013; Tavakoli et al. 2011). This current study aims to present a detailed and consolidated comparison of particle focusing behavior in the air-particle system with the conventional liquid-particle system. The effects of channel Reynolds number, particle's initial position, particle's diameter. and particle's density are considered.

### 3. Numerical methods

The numerical simulations are carried out by in-house CFD code based on the immersed boundary coupling with lattice Boltzmann method (IB-LBM). The lattice Boltzmann method (Aidun and Clausen 2009; Chen and Doolen 1998) has been a popular CFD numerical tool because of its advantages in simplicity, meshless, efficiency, and parallel scalability. Ladd (Chun and Ladd 2006; Ladd 1994a; Ladd 1994b) is the first one who employed the LBM for the simulation of particle migration behaviors in straight channels and calculated the interaction forces. And it has been used by several groups in the research of liquid-particle system in microfluidics (Asmolov et al. 2018; Haddadi and Di Carlo 2017; Wu et al. 2016). Recently, immersed boundary method (IBM) (Mittal and Iaccarino 2005) was further coupled with the LBM to overcome some shortcomings, such as calculation complexity of distinguishing interface between particle boundary and fluid, particle boundary unsmooth and pseudo oscillation of hydrodynamic force. Sun et al. (Dongke et al. 2013; Jiang et al. 2016; Sun and Bo 2015; Sun et al. 2016) have successfully applied this method to the particle-fluid interaction and complex particle dynamics in fluid-particle systems, including the moving, rotation, migration and focusing of a single particle or particle group in straight and non-straight microchannels.

In the IB-LBM coupled method, the fluid flow is solved by LBM based on fixed Cartesian orthogonal lattices and the particle is represented by a large number of moving Lagrangian points uniformly distributed at its surface. The particle-fluid hydrodynamic interaction is solved by the communication between the fixed fluid lattices and the moving surface points at the local region near the immersed particle surface (boundary). The numerical methods are briefly described below.

### 3.1. Immersed boundary-lattice Boltzmann method

For incompressible viscous flows laden with particles in this case, we adopt the single-relaxation time (SRT) LBGK model with the external force term (Guo et al. 2002b) as

$$f_\alpha(\mathbf{x}+\mathbf{e}_\alpha \Delta t, t+\Delta t) = f_\alpha(\mathbf{x},t) - \frac{1}{\tau}\left[f_\alpha(\mathbf{x},t) - f_\alpha^{(eq)}(\mathbf{x},t)\right] + F_\alpha(\mathbf{x},t)\Delta t, \tag{3}$$

where $f_\alpha$ is the velocity distribution function of a fluid particle moving in the α-th direction. In this study, 19 discrete directions of the D3Q19 model for 3D flows are adopted. $e_\alpha$, and $w_\alpha$ are the discrete velocity and the corresponding weighting coefficients, as $w_0=1/3$, $w_{1\sim 6}=1/18$, and $w_{7\sim 18}=1/36$. The discrete velocity vectors $e_\alpha$ are given by

$$[e_\alpha, \alpha=0,\cdots,18] = c\begin{bmatrix} 0 & 1 & -1 & 0 & 0 & 0 & 0 & 1 & -1 & 1 & -1 & 1 & -1 & 1 & -1 & 0 & 0 & 0 & 0 \\ 0 & 0 & 0 & 1 & -1 & 0 & 0 & 1 & -1 & -1 & 1 & 0 & 0 & 0 & 0 & 1 & -1 & 1 & -1 \\ 0 & 0 & 0 & 0 & 0 & 1 & -1 & 0 & 0 & 0 & 0 & 1 & -1 & -1 & 1 & 1 & -1 & -1 & 1 \end{bmatrix}, \tag{4}$$

where the lattice speed $c = \Delta x/\Delta t$, and $\Delta x$ and $\Delta t$ are, respectively, the lattice spacing step and time step. $f_\alpha^{(eq)}$ is the equilibrium distribution function with a form of

$$f_\alpha^{(eq)} = \omega_\alpha \rho_f \left[1 + \frac{3}{c^2}(\mathbf{e}_\alpha \cdot \mathbf{u}) + \frac{9}{2c^4}(\mathbf{e}_\alpha \cdot \mathbf{u})^2 - \frac{3}{2c^2}\mathbf{u}^2\right], \tag{5}$$

with $\mathbf{u}$ being the fluid velocity. The last term $F_\alpha(\mathbf{x},t)$ at the right hand of equation (3) is the external force term, defined as

$$F_\alpha(\mathbf{x},t) = \left(1 - \frac{1}{2\tau}\right)\omega_\alpha\left[3\frac{\mathbf{e}_\alpha - \mathbf{u}(\mathbf{x},t)}{c^2} + 9\frac{\mathbf{e}_\alpha \cdot \mathbf{u}(\mathbf{x},t)}{c^4}\mathbf{e}_\alpha\right] \cdot \mathbf{f}(\mathbf{x},t), \tag{6}$$

where $\mathbf{f}(\mathbf{x}, t)$ is the local force term acting on the fluid calculated by immersed boundary method. $\tau$ is the dimensionless relaxation time, determined by $\tau = 0.5 + 3\nu/(c^2\Delta t)$. The fluid density $\rho_f$ and velocity $\mathbf{u}$ corrected by the contribution from the immersed boundary force density $\mathbf{f}(\mathbf{x},t)$ are as follows:

$$\rho = \sum_\alpha f_\alpha \tag{7}$$

$$\mathbf{u} = \frac{1}{\rho}\sum_\alpha \mathbf{e}_\alpha f_\alpha + \frac{\Delta t}{2\rho}\mathbf{f}(\mathbf{x},t). \tag{8}$$

The immersed boundary method is usually implemented in three main steps (Cao et al. 2015; Dupuis et al. 2008; Jiang and Liu 2018):

**Velocity Interpolation:** the first step is to solve the intermediate fluid velocity $u_{bn}$ at the particle surface by interpolation the fluid velocity $u_f$ from the nearby Eulerian lattices $(x_f, y_f, z_f)$ to the Lagrangian points $(X_n,$

$Y_n$, $Z_n$), as

$$u_{bn} = \sum_f u_f \frac{1}{\Delta x^3} \delta\left(\frac{x_f - X_n}{\Delta x}\right) \delta\left(\frac{y_f - Y_n}{\Delta x}\right) \delta\left(\frac{z_f - Z_n}{\Delta x}\right) \cdot \Delta x^3. \tag{9}$$

**Boundary force calculation:** then calculate the IB-related force density at the Lagrangian points according to the no-slip boundary condition:

$$F_{bn} = \frac{2\rho_f}{\Delta t}(U_{pn} - u_{bn}). \tag{10}$$

**Boundary force spreading:** the third step is spreading the obtained force density $F_{bn}$ from the Lagrangian points to the neighboring Eulerian lattices:

$$f_f = \sum_n F_{bn} \frac{1}{\Delta x^3} \delta\left(\frac{x_f - X_n}{\Delta x}\right) \delta\left(\frac{y_f - Y_n}{\Delta x}\right) \delta\left(\frac{z_f - Z_n}{\Delta x}\right) \cdot (drs_n \cdot dA_n), \tag{11}$$

where $drs_n$ and $dA_n$ are the boundary thickness and the surface area of the surface point $n$. This obtained boundary force density $f_f$ can be substituted to the right hand of the Equations (6) and (8) to correct the fluid flow. In the interpolation and spreading steps according to Equations (9) and (11), the regularized Dirac delta function $\delta_{ij}(\cdot)$ is used, due to its smooth, compact support and calculation efficiency. Here we use the 4-points functions corresponding to the maximum scope of influence being 4 lattices for each boundary Lagrangian point (Jiang and Liu 2018; Peskin 2002) as:

$$\delta(r) = \begin{cases} 0, & |r| \geq 2, \\ \frac{1}{8}\left(5 - 2|r| - \sqrt{-7 + 12|r| - 4|r|^2}\right), & 1 \leq |r| < 2, \\ \frac{1}{8}\left(3 - 2|r| + \sqrt{1 + 4|r| - 4|r|^2}\right), & 0 \leq |r| < 1 \end{cases} \tag{12}$$

Based on this, the hydrodynamic force $F_h$ and moment $T_h$ acting on the particle can be directly calculated by integration of the surface forces and moments as:

$$F_h = \sum_n F_{bn} \cdot (drs_n \cdot dA_n), \tag{13}$$

$$T_h = \sum_n F_{bn} \cdot (drs_n \cdot dA_n) \times r_n, \tag{14}$$

with $r_n$ being the arm of the force.

### 3.2. Governing equations for particle motion

The governing equations for the translational and rotational motions of particles using the Newton's second law are as follows:

$$m_p a_p^{t+\Delta t} = F_h + F_b + F_{p-p/w} \tag{15}$$

$$\frac{u_p^{t+\Delta t} - u_p^t}{\Delta t} = 0.5\left(a_p^{t+\Delta t} + a_p^t\right) \tag{16}$$

$$\frac{x_p^{t+\Delta t} - x_p^t}{\Delta t} = 0.5\left(u_p^{t+\Delta t} + u_p^t\right) \tag{17}$$

$$I_\mathrm{p}\alpha_\mathrm{p}^{t+\Delta t} = T_h \tag{18}$$

$$\frac{\omega_\mathrm{p}^{t+\Delta t} - \omega_\mathrm{p}^t}{\Delta t} = 0.5\left(\alpha_\mathrm{p}^{t+\Delta t} + \alpha_\mathrm{p}^t\right) \tag{19}$$

$$\frac{\theta_\mathrm{p}^{t+\Delta t} - \theta_\mathrm{p}^t}{\Delta t} = 0.5\left(\omega_\mathrm{p}^{t+\Delta t} + \omega_\mathrm{p}^t\right) \tag{20}$$

where $m_p$ and $I_\mathrm{p}$ are, respectively, the particle's mass and moment inertia. $a_\mathrm{p}$, $u_\mathrm{p}$, $\alpha_\mathrm{p}$, $w_\mathrm{p}$, and $\theta_\mathrm{p}$ are the translation acceleration and velocity, rotational acceleration, velocity and angular of the particle $p$, respectively. The second term of the right hand of Equation (15) represents the buoyant force as $F_\mathrm{b} = (\rho_\mathrm{f}/\rho_\mathrm{p} -1.0) \cdot m_p g$, which is constant due to the constant densities of fluid and particle. The term $F_\mathrm{p-p/w}$ is the short range interaction force between particle and particle/wall, which can be calculated by lubrication model when the net distance between the particle surface and the channel wall is less than a lattice (Nguyen and Ladd 2002).

### 3.3. Code validation

To validate the accuracy of our in-house code based on the IB-LBM, two benchmark cases are implemented. The first one is the lateral migration of a freely moving and neutrally buoyant particle in a 3D square microchannel filled with liquid at $Re_\mathrm{c}$=100, referred to the previous literature (Lashgari et al. 2017). The geometry and the particle migration trajectories in the channel cross-section are shown in Fig. S1(a)-(b) in the supplementary file. Two obvious migration steps are observed: the first fast migration from the initial position to the equilibrium plane parallel to the wall and the second slow migration to the final equilibrium position along the isoplane of the streamwise velocity. In addition, the particles from different initial positions except the symmetric planes migrate to the same equilibrium position (approximately $0.2H$ away from the channel wall). The qualitative migration processes and quantitative equilibrium position both agree well with the literature (Lashgari et al. 2017). In the second case, we calculate the lateral lift force coefficient $f_\mathrm{L} = F_\mathrm{z} / (\rho_\mathrm{f} U_\mathrm{f}^3 / H)$ for further quantitative comparison, which has been obtained by (Di Carlo et al. 2009) and (Nakagawa et al. 2015) with finite element method and immersed boundary method, respectively. The $F_\mathrm{z}$ is the lateral lift force on the particle and can be solved by Equation (13). In this case, the particle is fixed in the $y$ and $z$ directions while freely moving in the streamwise $x$ direction and freely rotating in any direction. The results are shown in Fig. S1(c) in the supplementary file. It can be seen that our results also agree well with the corresponding results of (Di Carlo et al. 2009) and (Nakagawa et al. 2015).

### 3.4. Simulation setup

In this work, a straight microchannel with square cross-section are used and scaled as $L\times H\times H = 90\times 20\times 20$ μm，as shown in Fig. 2(a). The Poiseuille flow in this microchannel is driven by the pressure difference $\Delta P$ between inlet and outlet. The hollow arrowhead depicts the flow direction. The fluid flow are characterized by an approximate theoretical average bulk velocity $U_\mathrm{f}$ and the corresponding channel Reynolds number defined as $U_\mathrm{f} = \Delta P H^2 / 32\mu_\mathrm{f} L$ and $Re_\mathrm{c} = \rho_\mathrm{f} U_\mathrm{f} H / \mu_\mathrm{f}$ (Sun et al. 2016), respectively. The inlet and outlet boundary conditions are set to be periodical. The velocity and pressure boundary conditions at the wall, inlet and outlet are implemented by non-equilibrium extrapolation method with second order accuracy (Guo et al. 2002a). The lattice grid step is set as $\Delta x$ = 0.3125 μm after grid independence test for the uniform and orthogonal lattices, as shown in Fig. 2(b). The number of the lattices is about 289×65×65 = 1.22 million. The resolution of particle diameter is $d_p / \Delta x$ = 16 for the PM5.0 particle and 790 Lagrangian points are uniformly distributed at its surface to represent the particle. The time step is set as $\Delta t$ = 2.5×10$^{-9}$ s to ensure that the relaxation time $\tau$ and Mach number $Ma$ can satisfy the as $0.5 < \tau < 1.0$ and $Ma \ll 1$ for accuracy and stability requirements. In addition, all the calculations are speedup by MPI parallel strategy on 12 computational cores.

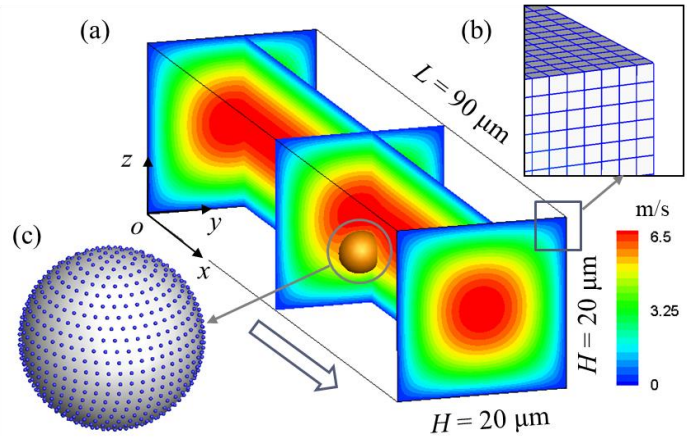

**Fig. 2.** Computational setup: (a) the streamwise velocity at different cross section of the microchannel when the PM5.0 particle reaches its equilibrium position at $Re_c$=2.0. (b) the computational fluid lattice used in the numerical simulation. (c) the Lagrangian points distributed on the particle surface.

## 4. Results and discussions
### 4.1. Interaction between fluid and particle

In the present work, the simulation is firstly carried out for PM5.0 particle at $Re_c$ = 2. Fig. 3 shows the air streamwise velocity, shear rate and shear rate gradient at the cross section of the straight microchannel. The air velocity, shear rate and the shear rate gradient all are significantly disturbed due to the presence of a finite-size PM5.0 particle. When particle is absent, they are symmetrical based on the horizontal midline, vertical midline and diagonal lines, as shown in Fig. 3(a)-(c). The maximum value of the velocity and the minimum value of the shear rate are at the center of the channel, while the minimum values of the shear rate gradient are positioned at the four diagonal lines. However, when particle is present, it occupies a certain volume and hence changes the distribution of flow field around the particle to be asymmetrical, as shown in Fig. 3(d)-(f). And also we can find that the shear rate and shear rate gradient are bigger in the vicinity of the particle. These phenomena are the results of the relative slip motion between particle and fluid caused by the big value of particle's Stokes number. And this will conversely have significant influence on the particle's motion. For example in Fig. 3(e), the shear rate at the bottom zone below the particle are much larger than that at the overhead zone above the particle. This will cause particle to rotate from up to bottom, with the direction as the red dashed arrow depicted. The rotation will generate the Magus force (Barkla and Auchterlonie 1971) to push the particle moving towards the channel center. However, it should be pointed out that this effect can't be obtained by the traditional Eulerian-Lagrangian method, in which particle is only treated as a Lagrangian point.

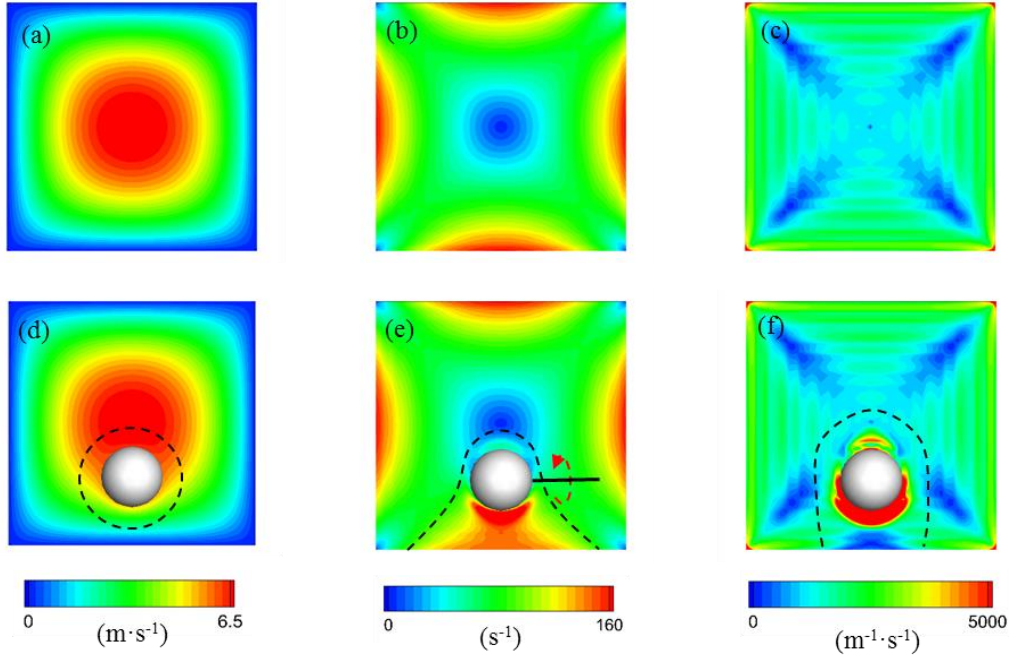

**Fig. 3.** Fluid streamwise velocity, shear rate and shear rate gradient with and without a particle in the straight microchannel at $Re_c=2$. (a), (b) and (c): Pure fluid without particle; (d), (e) and (f): Fluid with a single PM5.0 particle. (a) & (d): Streamwise velocity; (b) & (e): Shear rate calculated based on the streamwise velocity; (c) & (f): Shear rate gradient calculated based on the streamwise velocity. The locally disturbed zones affected by the particle are approximately depicted by the enclosed dashed black curves. The dashed red curve line in (e) represents the particle rotational direction as from paper to outside and then downside.

Fig. 4 shows the middle *x-z* plane view of the fluid pressure $P^*$, streamwise velocity $U$ and vertical velocity $W$ in the presence of a single particle located at non-symmetrical position $Z^*=0.85$ (Fig. 4(a)-(c)) and symmetrical position $Z^*=0.5$ (Fig. 4(d)-(f)). The pressure $P^*$ has been dimensionless based on $P_0$ as $P^* = P / P_0$. And the particle's coordinates have been dimensionless based on the channel height as $X^* = x_p / H$, $Y^* = y_p / H$ and $Z^* = z_p / H$. The pressure strip from left to right can be obviously observed in Fig. 4(a) and (d) because the fluid is driven by the pressure difference $\Delta P$ between left inlet with $P_{in} = P_0 + 0.5\Delta P$ and the right outlet with $P_{out} = P_0 - 0.5\Delta P$. And also we can find that the flow field is significantly disturbed by the particle though only at the local positions. Due to the interaction between particle and channel wall, we can find that the fluid pressure is low at left and high at right over the particle, while high at left and low at right below the particle, as shown in the Fig. 4(a). This will promote on the particle's surface from left to right at the lower part of the particle and from right to left at the upper part of the particle, and hence generate a particle rotation with a counter-clockwise direction as depicted by the black curved arrow in Fig. 4(a). Conversely, the particle counter-clockwise rotation will induce fluid flowing upward at right and downward at left in Fig. 4(c) because the non-slip boundary condition should be satisfied. However, when the particle is located at the central plane between the top and bottom walls, both the fluid pressure and velocity are symmetric about the particle's horizontal axis, as shown in Fig. 4(d)-(f). The symmetrical distribution of pressure cannot induce particle rotating.

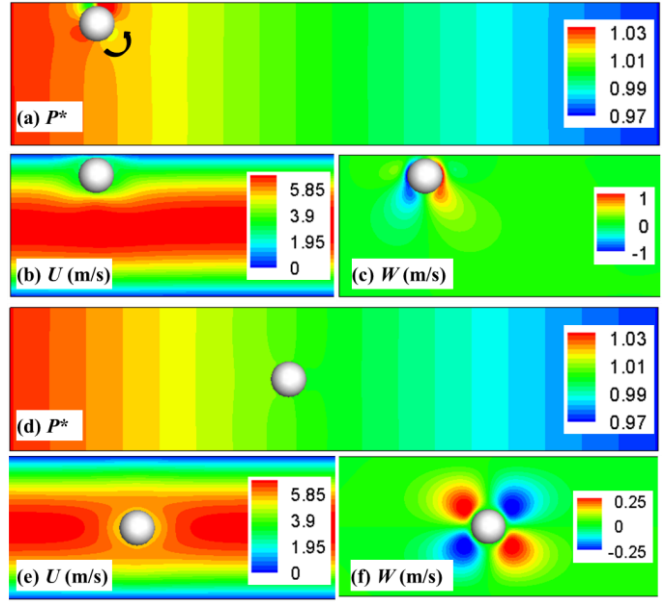

**Fig. 4.** Middle *x-z* plane view of the contour plot of the fluid dimensionless pressure $P^*$ (= $P / P_0$), streamwise *x*-velocity $U$ and vertical *z*-velocity $W$ in the microchannel for particle's location at $Z^*$=0.85: (a)-(c) and $Z^*$=0.5: (d)-(f) at $Y^*$=0.5 and $Re_c$=2.

Furthermore, the pressure distribution at the particle's surface is obtained by interpolation calculation as shown in Fig. 5. The dimensionless pressure coefficient is defined as $C_p = (P - P_0)/(0.5\rho_f U^2)$. The pressure coefficient at the position closing to the wall is relatively larger than that located at the channel center. It is symmetrical about the horizontal axis, large at the upstream surface and small at the downstream surface, as shown in Fig. 5(b) and (d). Therefore, the particle can be pushed by the pressure difference at its surface and translate horizontally with no rotation. However, the distribution of the pressure coefficient is non-uniform when the particle is near the channel wall. As can be seen in Fig. 5(a) and (c), the pressure difference between upstream surface and downstream surface is negative on the upper part while positive on the lower part. Hence, the generated pressure forces $F_1$ and $F_2$ which can produce moment about the particle's center with arm $L_1$ and $L_2$, ultimately yielding particle rotation in the counter-clockwise direction.

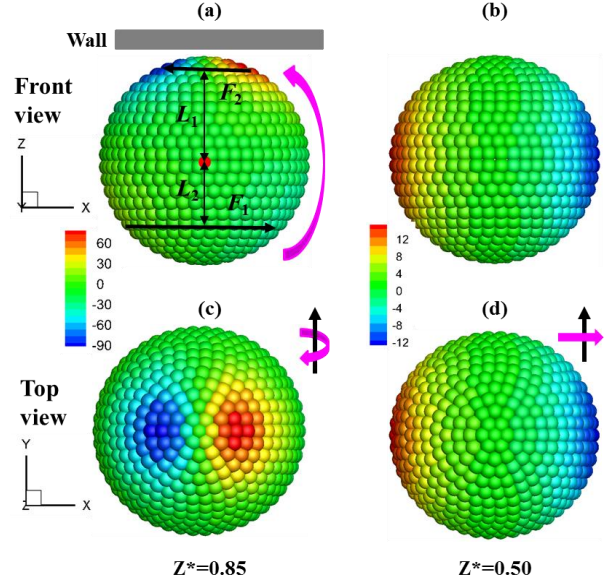

**Fig. 5.** The contour plot of the dimensionless pressure coefficient at the particle surface with its center located at Z*=0.85 (a, c) and Z*=0.50 (b, d) at $Re_c$=2.0 and $\rho_p^*$=830. Red curved arrows depict the rotational direction of the particle.

### 4.2. Particle's motion and focusing

We fix particle at $y$ and $z$ coordinates and it translates in $x$ direction with free rotation. Fig. 6(a) and (b) show the streamwise velocity $u_p$ and dimensionless angular velocity $\omega_p^*$ ($= 0.5 d_p \omega_p / U_f$) of the PM5.0 particle for different particle density ratios, $\rho_p^*$ ($= \rho_p / \rho_f$). The particle's velocities are lower than the fluid velocities at the same positions, as shown in Fig. 6(a). Particles with high density ratio, i.e. heavier particles, have lower velocities. It suggests that these aerosol particles lag behind fluid, and heavier particles lag behind even more, which are the source of the Saffman force described in Section 2. Interestingly, we find that the dimensionless angular velocity $\omega_p^*$ varies linearly with the particle position $Z^*$, and is not affected by the particle density (Fig. 6(b)). The value of the dimensionless angular velocity $\omega_p^*$ is positive when the particle is at the lower part ($Z^* < 0.5$), while negative when the particle is at the upper part ($Z^* > 0.5$). This means the passive particle driven by the Poseuille flow always rotates in a direction from channel center forward to channel wall and experiences Magnus force towards channel center, as shown in Fig. 6(c). When it travels through channel center from upward to downward the direction of the rotation and corresponding Magnus force will reverse.

The lateral lift force coefficient $f_L$ is also obtained as shown in Fig. 6(d). We can find that the lift force coefficient of the aerosol particle has the same trend as that of the neutrally buoyant particle in the liquid-particle system (in Fig. S1(c)). The points A and B corresponding to zero value of $f_L$ are the lower and upper equilibrium positions. And the heavier particle prefer equilibrium slightly closer to the channel center. Point C may be also an equilibrium position, which is higher than the channel center due to the balance of the lift force to the negative gravity. However, we think that it is an unstable equilibrium position like that in the liquid-particle system (Abbas et al. 2014; Nakagawa et al. 2015).

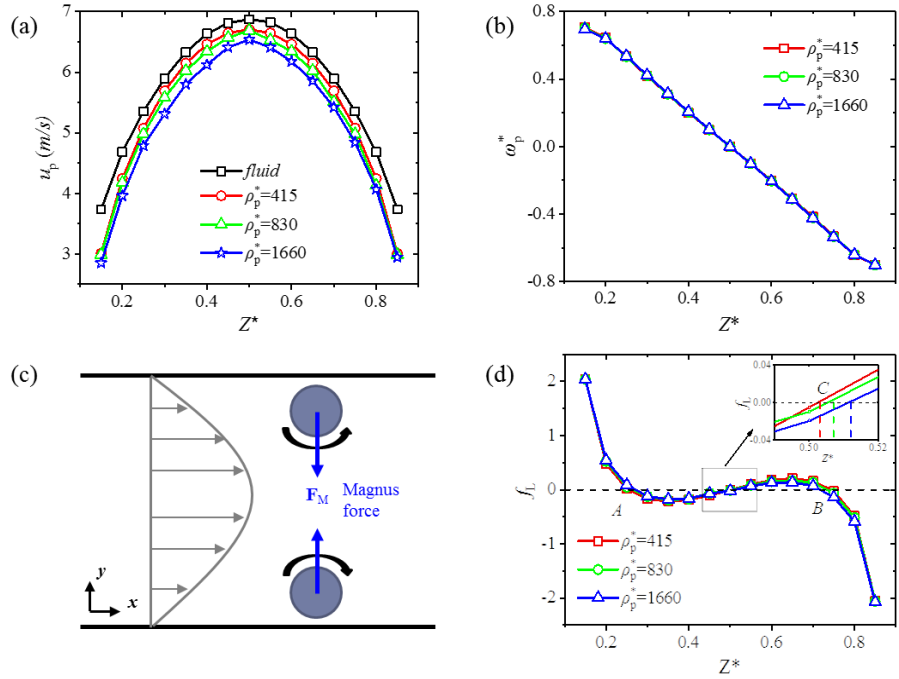

**Fig. 6.** Dimensionless streamwise velocity, angular velocity and lateral lift force coefficient of a single particle with different densities located at different $Z^*$ positions and at $Re_c$=2.

Next, the particle is placed at different initial positions $Z^*$ at the fixed symmetric and vertical midline $Y^*$=0.5 to freely move after the fluid flow is calculated to be stable. Fig. 7 shows the trajectories of a single particle. Particles finally reach two stable equilibrium positions (upper $Z_1^*$=0.756 and lower $Z_2^*$=0.241) after experiencing an overshoot. This overshoot is due to the large relative velocity difference between fluid and particle at the beginning, which generates large Saffman force (Fan et al. 1992) driving particles moving towards the channel center. However, these overshoots are finally eliminated by other lateral lift forces. Due to the gravity effect, the particles initially close to the channel center prefer migrating to the lower equilibrium position. In addition, the upper equilibrium position is slightly unsymmetrical with the lower one due to gravity effect.

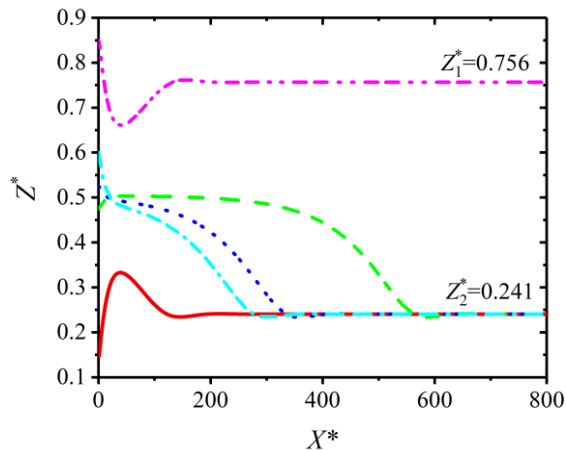

**Fig. 7.** Trajectories of a single PM 5.0 particle at $Re_c$ = 2 and $\rho_p^*$ = 830.

Furthermore, we found that the equilibrium positions in the air-particle system are different from those in the liquid-particle system. In a liquid-particle system, generally there are 9 equilibrium points (Abbas et al. 2014; Nakagawa et al. 2015; Zhang et al. 2016): 4 stable points at the vertical and horizontal midlines, 4 unstable points at the diagonal lines and 1 unstable point at the channel center point, as shown in Fig. 8(a). They are all symmetrical in both horizontal and vertical directions. However, in the air-particle system there are only 6 equilibrium points (1-6), as shown in Fig. 8(b). They are symmetrical only in the horizontal direction, i.e. left to right, by the vertical middle plane ($Y^*=0.5$), while un-symmetrical in the vertical direction, i.e. the gravity direction or up to bottom. In addition, the two equilibrium points (1-2) at the vertical midlines are unstable, while the other four equilibrium points (3-6) at the vicinity of the diagonal lines are stable. This stability phenomenon will be further confirmed in section 4.3 and however, are contrary to those in the liquid-particle system as mentioned above.

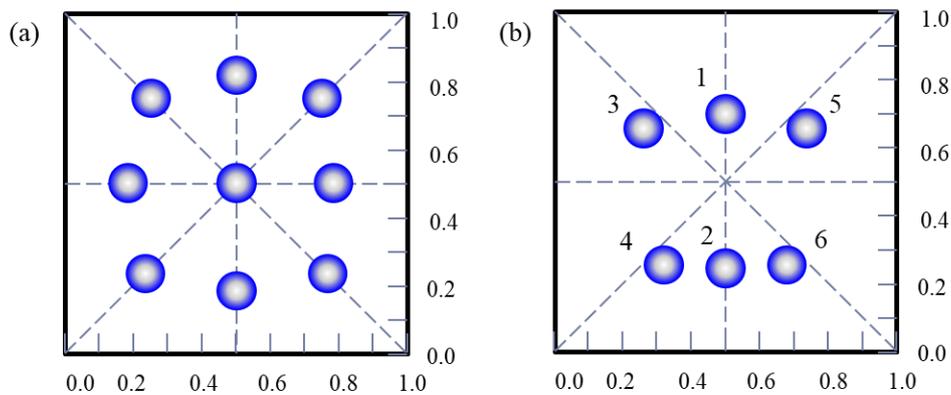

**Fig. 8.** Equilibrium positions of particles in a square straight microchannel in (a) liquid-particle system and (b) air-particle system.

### 4.3. Effects of the initial particle position

We also found that different initial particle positions significantly influence the particle's path and final equilibrium position. This is because the velocity gradient and the relative velocity at different positions are different, yielding different values of the lift forces. Fig. 9(a) and (b) show the particle movement with initial positions at the midlines and diagonal lines, and the inner zone. The hollow circles represent the initial positions of the particles. We can clearly observe six different equilibrium points, as shown by the solid circles. In the liquid-particle system particle always migrates to the nearby equilibrium point located in the same 1/8 zone. In other words the initial position and the final equilibrium position of the same particle are in the same 1/8 triangle zone, as can be shown in Fig. S1(c) and previous literatures (Abbas et al. 2014; Chun and Ladd 2006; Nakagawa et al. 2015). However, in Fig. 9 we can see that some particles have initial and final equilibrium positions positioned at different 1/8 triangle zones. In addition, only the particle initially located in the vertical midline is still focused at the vertical midline. All other particles even ones very close to the vertical midline migrate to the equilibrium positions below the diagonal lines. This indicates that the two equilibrium positions at the vertical midlines are not stable since a particle slightly disturbed away from the vertical midline will not migrate to the two equilibrium positions. The other four equilibrium positions at the vicinity of the diagonal lines are stable because particles finally migrate to these positions as long as the initial positions are not at the vertical midlines, as shown in Fig. 9(a) and (b).

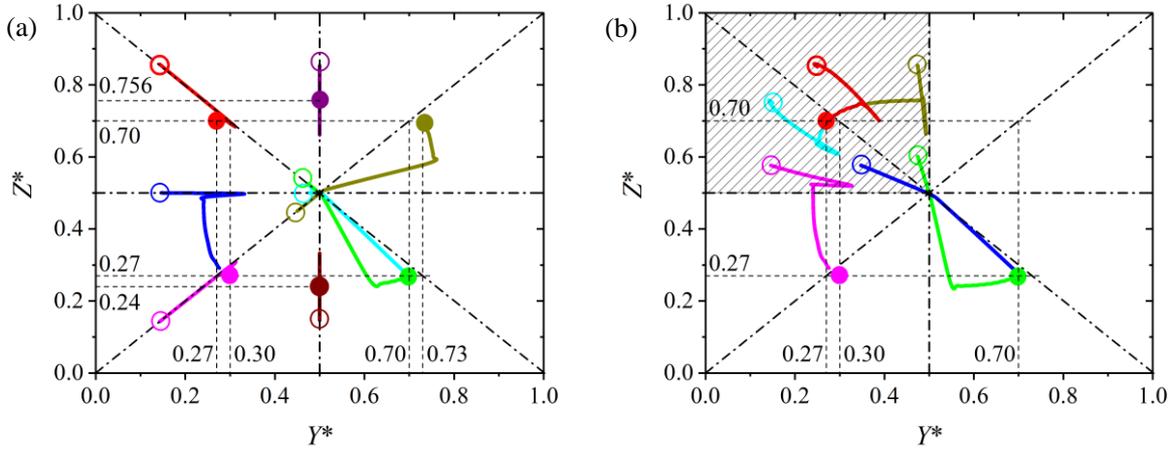

**Fig. 9.** Migration trajectories affected by the different initial particle positions. (a) Initial positions at the midlines and diagonal lines. (b) Initial positions not at the midlines or diagonal lines. The hollow circles and the solid circles represent the initial positions and the final equilibrium positions of the particles, respectively.

**4.4. Effects of the particle's density**

Fig. 10(a) and (b) illustrate the particle's streamwise trajectories affected by different particle densities. They show that particles with a higher density have a larger overshoot and longer streamwise distance to reach their equilibrium positions. This is because heavier particles have higher inertia due to their larger Stokes number depending on density $\rho_p$. However, like the lift force coefficient discussed in Section 4.2, the particle's final equilibrium positions are not affected obviously by the particle's density except selecting either the upper one ($Z^* \approx 0.754$) or the lower one ($Z^* \approx 0.241$). When particle have a lower density, it behaves approximately like the neutrally buoyant particle in the liquid-particle system, directly migrating to the nearest equilibrium position. The overshoot increases as the particle density increase. If the particle across the horizontal midline to enter another local zone, it will never return to its original equilibrium position and finally reach another equilibrium position.

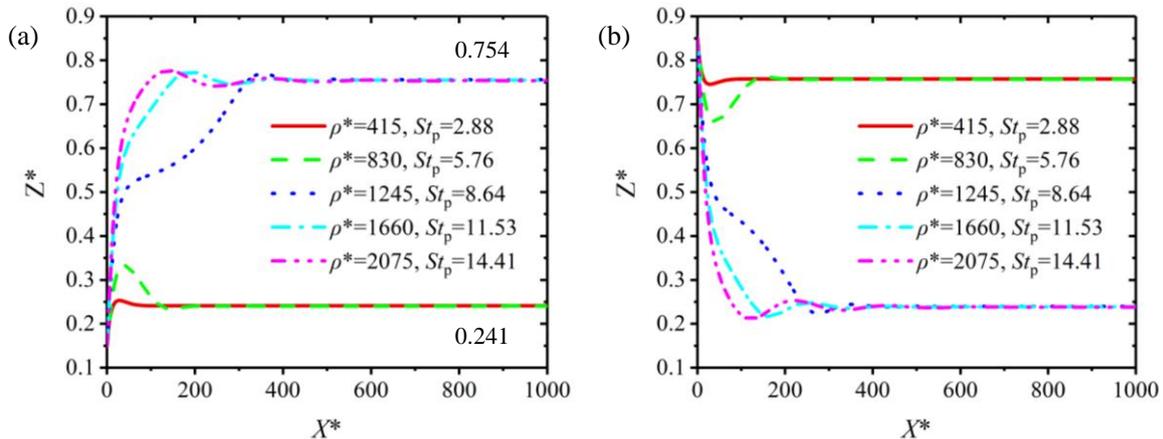

**Fig. 10.** Particle streamwise path with various particle densities at $Re_c = 2$. (a) initial position is at $Z^*=0.15$ and (b) initial position is at $Z^*=0.85$.

### 4.5. Effects of the particle's diameter

Lateral forces acting on the particle originate from the fluid pressure and velocity around the particle. These parameters change as the diameter of the particle varies. Fig. 11 presents the effect of particle's diameter on the fluid streamwise velocity at the cross section corresponding to the particle's center when the particle reaches its equilibrium position. The surface color of the particle depicts its $x$-velocity. It can be observed that larger particle occupies larger volume and has lower velocity, hence disturbs the fluid flow more significantly. And intuitively larger particle migrates closer to the channel's wall.

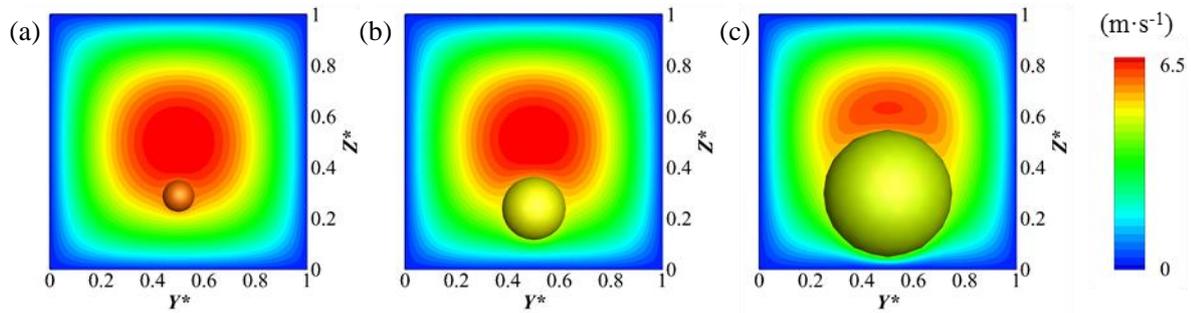

**Fig. 11.** Effect of the particle's diameter on the equilibrium position at $Re_c = 2$ and $\rho_p^* = 830$. (a) $D_p = 2.5$ μm, (b) $D_p = 5$ μm and (c) $D_p = 10$ μm.

Here, we define a parameter to investigate the effect of the particle's diameter on the equilibrium position by the ratio of the net distance between the particle's surface and the channel wall to the channel height as

$$\bar{Z}^* = (Z_p - 0.5 D_p)/H. \tag{21}$$

According to the definition, $\bar{Z}^*$ is a dimensionless quantity that increases from zero as the particle moves from the bottom wall to the top wall. Fig. 12 shows the migration process and the final equilibrium position of particles with different diameters. Small particle ($D_p = 2.5$ μm) focuses faster and has the tendency to follow the flow direction. While larger particles migrate to lower positions with an oscillation.

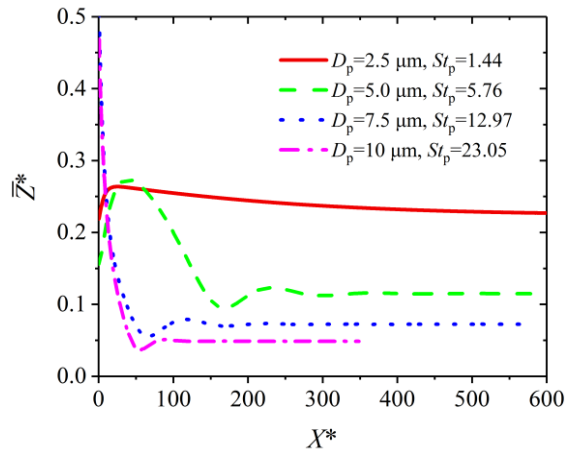

**Fig. 12.** Effect of particle's diameter on the streamwise path at $Re_c = 2$ and $\rho_p^* = 830$.

### 4.6. Effects of the fluid Reynolds number

At last, the channel Reynolds number $Re_c$ is varied, as shown in Fig. 13. We can see that larger $Re$ number will cause larger overshoot. This because the fluid shear rate increase with the increase in $Re_c$. And an interesting phenomenon can be found that the particle path is oscillating about the final equilibrium position when the Reynolds number is larger than 2. The oscillation pattern is like sine function. And its magnitude and duration time increase as the Reynolds number increases. This may because the Saffman force has been dominate when the Reynolds number is high. This lift force is proportional to the shear rate and the magnitude of the relative velocity between the particle and fluid as (Saffman 1965):

$$F_{Sa} = 1.615 D_p^2 (u_f - u_p) \sqrt{v_f \left| \frac{\partial u_f}{\partial z} \right|} \, \text{sign}\left( \frac{\partial u_f}{\partial z} \right), \tag{22}$$

where $(u_f - u_p)$ is the relative axial velocity component of the particle with respect to the fluid and $\partial u_f / \partial z$ is the shear rate of the axial velocity component at the vertical direction. The shear rate and relative velocity are relevant to the particle's position $Z^*$. The competition between this Saffman force with other lift force cause the oscillation phenomenon.

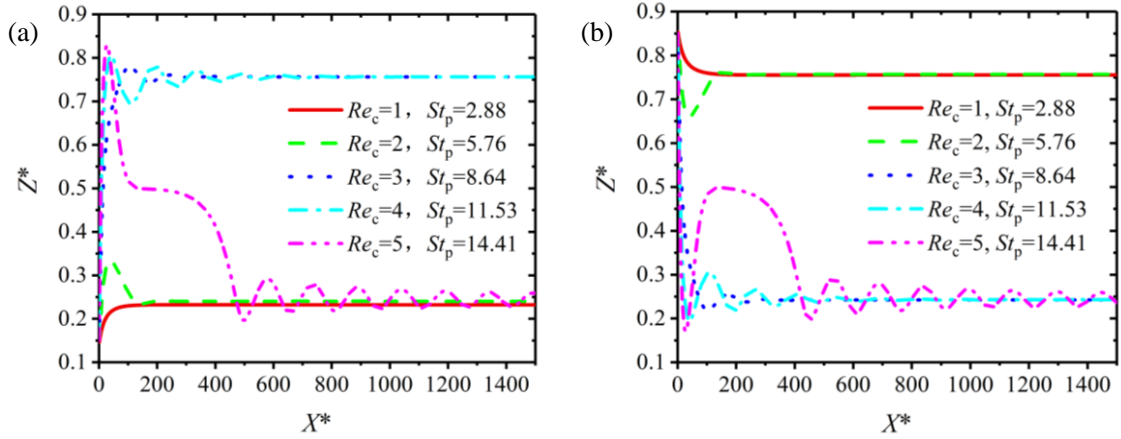

**Fig. 13.** Effect of the channel Reynolds number $Re_c$ on the streamwise path at $\rho_p^* = 830$ and $D_p = 5$ μm. (a) initial position is at $Z^*=0.15$ and (b) initial position is at $Z^*=0.85$.

### 5. Conclusions

Monodisperse aerosol particles' motions in rectangular microchannels are investigated by IB-LBM simulations to observe their inertial migration behavior. The initial position, diameter, density of the particle and the fluid Reynolds number are found to be the critical parameters. Conclusions are listed as follows:

1) The forces the particle experienced in the air-particle system are more complex than those in the liquid-particle system. The additional Saffman and Mangus forces will dominate when the fluid Reynolds number is high.

2) Similar to the liquid-particle system, there are two steps during the particle migration process in the air-particle system. First is the fast migration through the streamwise velocity contour and the second is the slow migration along the streamwise velocity contour towards the closest equilibrium position.
3) There are six equilibrium positions. Two unstable equilibrium positions are located at the vertical midlines, while four stable equilibrium positions locate beneath the diagonal lines. They are symmetrical in horizontal direction while unsymmetrical in vertical direction due to gravity effect.
4) Particles have an overshoot prior to reach their equilibrium positions, and the degree of this overshoot increases as the particle's diameter and density increase.
5) The initial position of the particle affects the migration trajectory and the final equilibrium positions.
6) High Reynolds number can cause particle oscillation near the final equilibrium position.

**Acknowledgments**

This work is supported by the National Natural Science Foundation of China (NSFC) (Grant Nos. 51876075, 51876076) and the Foundation of State Key Laboratory of Coal Combustion (Grant No. FSKLCCA1802).

**References**

Abbas M, Magaud P, Gao Y, Geoffroy S (2014) Migration of finite sized particles in a laminar square channel flow from low to high Reynolds numbers Physics of Fluids 26:123301 doi:10.1063/1.4902952

Aidun CK, Clausen JR (2009) Lattice-Boltzmann Method for Complex Flows Annual Review of Fluid Mechanics 42:439-472 doi:10.1146/annurev-fluid-121108-145519

Akhatov IS, Hoey JM, Swenson OF, Schulz DL (2008a) Aerosol flow through a long micro-capillary: collimated aerosol beam Microfluidics & Nanofluidics 5:215-224

Akhatov IS, Hoey JM, Swenson OF, Schulz DL (2008b) Aerosol focusing in micro-capillaries: Theory and experiment Journal of Aerosol Science 39:691-709

Amini H, Lee W, Di Carlo D (2014) Inertial microfluidic physics Lab on a Chip 14:2739-2761 doi:10.1039/C4LC00128A

Amini H, Sollier E, Weaver WM, Carlo DD (2012) Intrinsic particle-induced lateral transport in microchannels Proceedings of the National Academy of Sciences of the United States of America 109:11593

Asmolov ES, Dubov AL, Nizkaya TV, Harting J, Vinogradova OI (2018) Inertial focusing of finite-size particles in microchannels Journal of Fluid Mechanics 840:613-630 doi:10.1017/jfm.2018.95

Barkla HM, Auchterlonie LJ (1971) The Magnus or Robins effect on rotating spheres Journal of Fluid Mechanics 47:437-447 doi:10.1017/S0022112071001150

Bian X et al. (2016) Microfluidic Air Sampler for Highly Efficient Bacterial Aerosol Collection and Identification Analytical Chemistry 88:11504-11512 doi:10.1021/acs.analchem.6b02708

Brennen CE (2005) Fundamentals of multiphase flow. Cambridge university press,

Bruus H (2008) Theoretical microfluidics. Oxford University Press,

Cao C, Chen S, Li J, Liu Z, Zha L, Bao S, Zheng C (2015) Simulating the interactions of two freely settling spherical particles in Newtonian fluid using lattice-Boltzmann method Applied Mathematics and Computation 250:533-551 doi:https://doi.org/10.1016/j.amc.2014.11.025

Chen S, Doolen GD (1998) Lattice Boltzmann method for fluid flows Annual Review of Fluid Mechanics 30:329-364 doi:10.1146/annurev.fluid.30.1.329

Chin CD et al. (2011) Microfluidics-based diagnostics of infectious diseases in the developing world Nature Medicine 17:1015 doi:10.1038/nm.2408

Chun B, Ladd AJC (2006) Inertial migration of neutrally buoyant particles in a square duct: An investigation of multiple equilibrium positions Physics of Fluids 18:031704 doi:10.1063/1.2176587

Deng R, Zhang X, Smith KA, Wormhoudt J, Lewis DK, Freedman A (2008) Focusing Particles with Diameters of 1 to 10 Microns into Beams at Atmospheric Pressure Aerosol Science and Technology 42:899-915 doi:10.1080/02786820802360674

Di Carlo D, Edd JF, Humphry KJ, Stone HA, Toner M (2009) Particle Segregation and Dynamics in Confined Flows Physical Review Letters 102:094503

Dongke S, Di J, Nan X, Ke C, Zhonghua N (2013) An Immersed Boundary-Lattice Boltzmann Simulation of Particle Hydrodynamic Focusing in a Straight Microchannel Chinphyslett 30:74702-074702

Douwes J, Thorne P, Pearce N, Heederik D (2003) Bioaerosol Health Effects and Exposure Assessment: Progress and Prospects The Annals of Occupational Hygiene 47:187-200 doi:10.1093/annhyg/meg032

Dupuis A, Chatelain P, Koumoutsakos P (2008) An immersed boundary–lattice-Boltzmann method for the simulation of the flow past an impulsively started cylinder Journal of Computational Physics 227:4486-4498 doi:https://doi.org/10.1016/j.jcp.2008.01.009

Fan B, McFarland AR, Anand NK (1992) Characterization of the aerosol particle lift force Journal of Aerosol Science 23:379-388 doi:https://doi.org/10.1016/0021-8502(92)90007-I

Guo Z-L, Zheng C-G, Shi B-C (2002a) Non-equilibrium extrapolation method for velocity and pressure boundary conditions

in the lattice Boltzmann method Chinese Physics 11:366

Guo Z, Zheng C, Shi B (2002b) Discrete lattice effects on the forcing term in the lattice Boltzmann method Physical Review E 65:046308

Haddadi H, Di Carlo D (2017) Inertial flow of a dilute suspension over cavities in a microchannel Journal of Fluid Mechanics 811:436-467 doi:10.1017/jfm.2016.709

Han E-T (2013) Loop-mediated isothermal amplification test for the molecular diagnosis of malaria Expert Review of Molecular Diagnostics 13:205-218 doi:10.1586/erm.12.144

Hoey JM, Lutfurakhmanov A, Schulz DL, Akhatov IS (2012) A Review on Aerosol-Based Direct-Write and Its Applications for Microelectronics Journal of Nanotechnology 2012:22 doi:10.1155/2012/324380

Hong SC, Kang JS, Lee JE, Kim SS, Jung JH (2015) Continuous aerosol size separator using inertial microfluidics and its application to airborne bacteria and viruses Lab on a Chip 15:1889-1897 doi:10.1039/C5LC00079C

Jebakumar AS, Premnath KN, Abraham J (2016) Lattice Boltzmann method simulations of Stokes number effects on particle trajectories in a wall-bounded flow Computers & Fluids 124:208-219 doi:https://doi.org/10.1016/j.compfluid.2015.07.020

Jiang D, Tang W, Xiang N, Ni Z (2016) Numerical simulation of particle focusing in a symmetrical serpentine microchannel RSC Advances 6:57647-57657 doi:10.1039/C6RA08374A

Jiang M, Liu Z (2018) A Boundary Thickening-based Direct Forcing Immersed Boundary Method for Fully Resolved Simulation of Particle-laden Flows arXiv preprint arXiv:180609403

Jing W, Jiang X, Zhao W, Liu S, Cheng X, Sui G (2014) Microfluidic Platform for Direct Capture and Analysis of Airborne Mycobacterium tuberculosis Analytical Chemistry 86:5815-5821 doi:10.1021/ac500578h

Jing W et al. (2013) Microfluidic Device for Efficient Airborne Bacteria Capture and Enrichment Analytical Chemistry 85:5255-5262 doi:10.1021/ac400590c

Ladd AJC (1994a) Numerical simulations of particulate suspensions via a discretized Boltzmann equation. Part 1. Theoretical foundation Journal of Fluid Mechanics 271:285-309 doi:10.1017/S0022112094001771

Ladd AJC (1994b) Numerical simulations of particulate suspensions via a discretized Boltzmann equation. Part 2. Numerical results Journal of Fluid Mechanics 271:311-339 doi:10.1017/S0022112094001783

Lashgari I, Ardekani MN, Banerjee I, Russom A, Brandt L (2017) Inertial migration of spherical and oblate particles in straight ducts Journal of Fluid Mechanics 819:540-561 doi:10.1017/jfm.2017.189

Lee K-S, Hwang T-H, Kim S-H, Kim SH, Lee D (2013) Numerical Simulations on Aerodynamic Focusing of Particles in a Wide Size Range of 30 nm–10 μm Aerosol Science and Technology 47:1001-1008 doi:10.1080/02786826.2013.808737

Liu C, Hu G, Jiang X, Sun J (2015) Inertial focusing of spherical particles in rectangular microchannels over a wide range of Reynolds numbers Lab on a Chip 15:1168-1177 doi:10.1039/C4LC01216J

Liu Q, Zhang Y, Jing W, Liu S, Zhang D, Sui G (2016) First airborne pathogen direct analysis system Analyst 141:1637-1640 doi:10.1039/C5AN02367J

Martel JM, Toner M (2014) Inertial Focusing in Microfluidics Annual Review of Biomedical Engineering 16:371-396 doi:10.1146/annurev-bioeng-121813-120704

Maxey M (2017) Simulation Methods for Particulate Flows and Concentrated Suspensions Annual Review of Fluid Mechanics 49:171-193 doi:10.1146/annurev-fluid-122414-034408

Metcalf AR, Narayan S, Dutcher CS (2018) A review of microfluidic concepts and applications for atmospheric aerosol science Aerosol Science and Technology 52:310-329 doi:10.1080/02786826.2017.1408952

Mittal R, Iaccarino G (2005) Immersed boundary methods Annual Review of Fluid Mechanics 37:239-261 doi:10.1146/annurev.fluid.37.061903.175743

Nakagawa N, Yabu T, Otomo R, Kase A, Makino M, Itano T, Sugihara-Seki M (2015) Inertial migration of a spherical

particle in laminar square channel flows from low to high Reynolds numbers Journal of Fluid Mechanics 779:776-793 doi:10.1017/jfm.2015.456

Nguyen NQ, Ladd AJC (2002) Lubrication corrections for lattice-Boltzmann simulations of particle suspensions Physical Review E 66:046708 doi:10.1103/PhysRevE.66.046708

Peskin CS (2002) The immersed boundary method Acta Numerica 11:479-517 doi:10.1017/S0962492902000077

Pietrowska M et al. (2009) Mass spectrometry-based serum proteome pattern analysis in molecular diagnostics of early stage breast cancer Journal of Translational Medicine 7:60 doi:10.1186/1479-5876-7-60

Saffman PG (1965) The lift on a small sphere in a slow shear flow Journal of Fluid Mechanics 22:385-400 doi:10.1017/S0022112065000824

Schaap A, Chu WC, Stoeber B (2012) Transport of airborne particles in straight and curved microchannels Physics of Fluids 24:083301 doi:10.1063/1.4742900

Sui G, Cheng X (2014) Microfluidics for detection of airborne pathogens: what challenges remain? Bioanalysis 6:5-7 doi:10.4155/bio.13.287

Sun D-K, Bo Z (2015) Numerical simulation of hydrodynamic focusing of particles in straight channel flows with the immersed boundary-lattice Boltzmann method International Journal of Heat and Mass Transfer 80:139-149 doi:https://doi.org/10.1016/j.ijheatmasstransfer.2014.08.070

Sun D-K, Wang Y, Dong A-P, Sun B-D (2016) A three-dimensional quantitative study on the hydrodynamic focusing of particles with the immersed boundary – Lattice Boltzmann method International Journal of Heat and Mass Transfer 94:306-315 doi:https://doi.org/10.1016/j.ijheatmasstransfer.2015.11.012

Tavakoli F, Mitra SK, Olfert JS (2011) Aerosol penetration in microchannels Journal of Aerosol Science 42:321-328 doi:https://doi.org/10.1016/j.jaerosci.2011.02.007

Wu Z, Chen Y, Wang M, Chung AJ (2016) Continuous inertial microparticle and blood cell separation in straight channels with local microstructures Lab on a Chip 16:532-542 doi:10.1039/C5LC01435B

Yin H, Wan H, Mason AJ Separation and electrochemical detection platform for portable individual PM2.5 monitoring. In: 2017 IEEE International Symposium on Circuits and Systems (ISCAS), 28-31 May 2017 2017. pp 1-4. doi:10.1109/ISCAS.2017.8050733

Yu F, Li Y, Li M, Tang L, He J-J (2017) DNAzyme-integrated plasmonic nanosensor for bacterial sample-to-answer detection Biosensors and Bioelectronics 89:880-885 doi:https://doi.org/10.1016/j.bios.2016.09.103

Zhang D, Bi H, Liu B, Qiao L (2018) Detection of Pathogenic Microorganisms by Microfluidics Based Analytical Methods Analytical chemistry

Zhang J, Yan S, Yuan D, Alici G, Nguyen N-T, Ebrahimi Warkiani M, Li W (2016) Fundamentals and applications of inertial microfluidics: a review Lab on a Chip 16:10-34 doi:10.1039/C5LC01159K